\documentclass[psfig,times]{aa}  
\input{psfig.sty}
\usepackage{times}           
\begin{document}
\def\Mr{${\rm M}_{\rm R}$ }
\def\Mi{${\rm M}_{\rm I}$ }
\def\mr{${\rm m}_{\rm R}$ }
\def\mi{${\rm m}_{\rm I}$ }
\def\re{${\rm r}_{\rm e}$ }
\def\p0537{PKS~0537$-$441}
\def\o2{$[$O II$]$}
\def\O3{$[$O III$]$}
\def\ha{H$\alpha$}
\def\hb{H$\beta$}
\def\mg2{Mg~II}
\title{
\p0537: extended \o2 emission and a binary QSO?
\thanks{Based on observations collected with the VLT$-$UT1 on Cerro Paranal 
(Chile) and the NTT on La Silla (Chile) operated by the European Southern
Observatory in the course of the observing proposals 64.P-0230 and 66.B-0125.}
}

\author{Jochen Heidt\inst{1}
\and
 Klaus J\"ager\inst{2}
\and
 Kari Nilsson\inst{3}
\and
Ulrich Hopp\inst{4}
\and 
Josef W. Fried\inst{5} 
\and
Eckhard Sutorius\inst{6}}
\offprints{J. Heidt}

\institute{Landessternwarte Heidelberg, K\"onigstuhl,
D$-$69117 Heidelberg, Germany
\and 
Universit\"atssternwarte G\"ottingen, Geismarlandstr. 11, D$-$37083
G\"ottingen, Germany   
\and 
Tuorla Observatory, FIN$-$21500 Piikki\"o, Finland
\and
Universit\"atssternwarte M\"unchen, Scheinerstr. 1, D$-$81679 M\"unchen,
Germany   
\and
Max-Planck-Institut f\"ur Astronomie, K\"onigstuhl 17, D$-$69117 Heidelberg,
Germany
\and
Royal Observatory Edinburgh, Blackford Hill, Edinburgh EH9 3HJ, United Kingdom
}

\date{ Received ... / Accepted ...}

\abstract{We present high-resolution imaging and low-resolution
spectroscopy of the BL Lac object
\p0537 (z = 0.893) and its environment carried out with the
  ESO-NTT and VLT telescopes. The observations were
  designed to clarify, whether the properties of \p0537 are affected by
  gravitational microlensing due to the claimed detection of a galaxy
  along the line-of-sight to the BL Lac, or whether \p0537 and its environment 
  act as a lensing system itself, as suggested by the detection of
  several closeby 
companion galaxies with similar morphologies close to \p0537.\\ 
Our observations show that neither case seems to be likely. Within our
  images we did not find a galaxy along the line-of-sight 
to the BL Lac as claimed
previously. 
In addition, our spectroscopy shows that none of the four closest
companion galaxies
(including one new detection by us)
is at high redshift. Instead, two of the four nearby
companion galaxies to \p0537 are within 200  km/s of the systemic 
velocity of the BL Lac (z = 0.892 and 0.895, respectively). 
The third companion galaxy is at higher redshift (z = 0.947). 
The fourth companion galaxy shows evidence of \mg2 absorption
redwards of its systemic velocity and is perhaps a mini
low-ionization BAL QSO at z = 0.885. 
If the latter can be confirmed, \p0537 is the
first BL Lacertae object being a member of a binary Quasar. \\
While we do not find (micro)lensing effects being important for
this source or its environment, our observations revealed a 
highly interesting feature. We detected extended \o2 emission in the
off-nuclear spectrum of \p0537, which is most likely due to
photoionization from the active nucleus, although we can not rule out
the possibility that the extended emission is due to jet-cloud
interaction with the counterjet of \p0537. \\
According to our analysis of the photometric data,
\p0537 seems to be located in a cluster environment as rich as Abell
type 0-1. This is supported by the detection of four galaxies in the
field with similar redshifts as the BL Lac ($\Delta$ z $<$ 0.002). 
However, we found
serendipitously even more galaxies at somewhat higher redshifts (z =
0.9-1). Thus, \p0537 might be located in front of a galaxy
cluster at somewhat higher redshift or even be part of a large-scale
structure with an extension towards the BL Lac.\\

}

\titlerunning{\p0537: extended \o2 emission and a binary QSO?}
\authorrunning{J. Heidt et al.}

\maketitle

\section{Introduction}

BL Lac objects are a relatively rare and extreme subclass of 
the Active Galactic Nuclei (AGN).
They are dominated by non-thermal emission from radio- up to
X-ray frequencies (in some cases even up to the $\gamma$-ray regime) 
with mostly weak or even absent emission lines in their optical spectra. 
Within the standard AGN picture, their non-thermal radiation emerges 
from a beamed, relativistic jet closely aligned along the line-of-sight 
to the observer (Urry \& Padovani \cite{urpa}). 

One of their defining characteristics is their rapid and strong variability
across the whole electromagnetic spectrum. There is common
agreement, that the observed variability is intrinsic to the sources
(see e.g. the review by Wagner \& Witzel \cite{sw}), 
although some BL Lac objects may indeed be micro-lensed background QSOs 
(Ostriker \& Vietri \cite{ostriker}).

According to the Unified Scheme for AGN, the parent population of 
BL Lac objects - i.e. those sources, whose jets are misaligned to the
observer - are the FR I radio galaxies (Urry \& Padovani \cite{urpa}). 
This can be tested by comparing their isotropic (unbeamed) properties 
(e.g. host galaxies or cluster environment), which should be very similar. 
Host galaxy studies allow us to test the micro-lensing scenario, 
since here an offset between the center of a detected galaxy 
and the AGN itself could be present. In this case the galaxy
is presumably a system along the line of sight 
rather than the host of the AGN. 

In recent years, in more than 100  BL Lac objects with 
z $<$ 0.5 a host galaxy has been 
detected (e.g. Falomo \cite{fal96}, 
Wurtz et al. \cite{wur96}, Heidt et al. \cite{hei99}, 
Falomo \& Kotilainen \cite{falkot}, Scarpa et al. \cite{scar00},
Nilsson et al. \cite{kari})
and there is common agreement that BL Lac host galaxies are luminous, 
giant elliptical galaxies. Host galaxy data for BL 
Lacs at higher redshifts are scarce, so any evolutionary trends can
not be investigated. 
The cluster environment has been studied by Wurtz et al.
(\cite{wur97}) for a sample of $\sim$ 50 sources up to z = 0.65 
(mostly at z $<$ 0.4). They found an apparent evolution towards denser
cluster environments with redshift. However, the apparent evolution is 
tied to  observations of  a few BL Lac objects between z = 0.5 and 0.65 in 
their sample only. Similar conclusions have been drawn by Fried et
al. (\cite{fried}) from a study of the 1 Jy sample of BL Lacs
(Stickel et al. \cite{sti93}). Unfortunately,
due to the limited  depth of their observations, they were not able to 
evaluate the cluster environment for high-redshift (z = 0.8-1) 1 Jy BL Lacs
(including \p0537).

Up to now, there is only one clear example of a lensed BL Lac (B2 0218+35.7,
Patnaik et al. \cite{pat93}). Other objects, whose properties may be affected
by gravitational (micro)lensing are AO 0235+164, \p0537 and
B2 1308+326 (see Heidt \cite{heirev} for details). These three BL Lacs
 all have similar properties. 
They are at relatively high redshift (z $\sim$ 1), are among the most
luminous BL Lac objects (\Mr $<$ -27.5) and have historically shown the 
largest variability amplitudes ($\Delta$ m $>$ 5 mag, see Tab. 1 in Stickel et
al. \cite{sti93}). A few more candidates detected during the course of the BL
Lac HST snap survey are discussed in Scarpa et al. (\cite{scarlens}).

Of particular interest is the BL Lac object \p0537 (z = 0.892, 
Lewis \& Ibata \cite{lew00}), which is a member of the 1 Jy sample of BL Lac
objects (Stickel et al. \cite{sti93}). It displayed intraday variability at 
radio (Romero et al. \cite{rom95}) and optical frequencies
(Heidt \& Wagner \cite{hei96}, Romero et al. \cite{rom00}), and 
is one of the few BL Lac objects,
where $\gamma$-ray emission has been detected (Hartman et al. \cite{har99}).
Its  continuum variations throughout the entire electromagnetic spectrum 
have been discussed by Pian et al. (\cite{pian02}). Interestingly, Romero et
al. (\cite{rom95}) derived from their radio variability measurements brightness
temperatures in excess of $10^{21}$ K, which is 9 orders of magnitude above
the inverse Compton limit. Based on their claimed detection of a galaxy along
the line-of-sight to \p0537 Stickel et al. (\cite{sti88}, see below) 
discussed, whether the BL Lac could be microlensed by small masses in this
foreground galaxy. On the other hand, spectral-index variations found from
optical two-color photometry by Romero et al. (\cite{rom00}) seem to rule
out the microlensing hypothesis at least for these observations.

\p0537 was also subject to several broad-band 
imaging studies through a R-filter, however, with puzzling
results. Stickel et al. (\cite{sti88}) claimed the detection of a
foreground disk galaxy to \p0537, with a morphology and 
brightness similar to a galaxy 11\arcsec east of the BL Lac at z = 0.186. 
They concluded that the observed properties of \p0537 are influenced by
gravitational lensing by the foreground galaxy.  In subsequent studies
Falomo et al. (\cite{fal92}) and Kotilainen et al. (\cite{kot98})
did not detect the foreground galaxy 
reported by Stickel et al. (\cite{sti88}) and found \p0537 
completely unresolved thus weakening the lensing hypothesis. 
The former noted two faint companion galaxies close to the BL Lac at $\sim$ 
4\arcsec distance. 
The BL Lac object also appeared unresolved in HST observations 
(Scarpa et al. \cite{scar00}). In a reanalysis of the same HST-image 
mentioned above, Lewis \& Ibata (\cite{lew00}) found indications for a
galactic component after the subtraction of a scaled PSF
(which was recently shown to be due to scattered light by Pian et al. 
\cite{pian02}). Additionally, they detected 3 companion galaxies 
close to \p0537 (2 of which had already been found by Falomo et al. 
\cite{fal92}) and discussed an alternative scenario in which \p0537 is 
not subject to lensing effects, but is rather a member of a group of galaxies
which lense a distant background galaxy. As a consequence at least two of the
three companion galaxies could be magnified and split images of a distant
background source at high redshift. 

Obviously, the results and the interpretation of the results for \p0537 differ 
grossly. One major drawback of these studies is that the images were either
taken under moderate seeing conditions or suffer from relatively short
exposure times.  Except for \p0537 itself and the nearby galaxy
11\arcsec to the east, no spectroscopy in the field of \p0537 has been
carried out until now. In order to clarify the situation, we have undertaken 
a very deep imaging and multi-object spectroscopic study of
\p0537 and its environment using the NTT and VLT telescopes at ESO, the
results of which are presented here. While we will show that
lensing effects are most likely not important in this source, several highly
interesting features are observed. Amongst them are the presence of extended
\o2 emission on the opposite side of the radio jet in \p0537 and the possible 
detection of a binary QSO.

To facilitate comparison with other studies, ${\rm H}_{\rm 0} =$ 50\ km\
${\rm s}^{\rm -1}\ {\rm Mpc}^{\rm -1}$ and ${\rm q}_{\rm 0} = 0$ is assumed.

\section{Observations and data reduction}

\subsection{Imaging}

A very deep I-band image of \p0537 was acquired at the NTT in 
service mode on the night November 7/8 1999. The use of the I-filter 
instead of a R-filter (which was used in previous studies) has the advantage, 
that it samples the spectrum redwards of the 4000\AA\ break at z = 0.893
thus increasing the chance to detect the host gfalaxy of \p0537.
We used SUSI2,
which consists of  a mosaic of two 2k$\times$4k EEV-CCDs. The frames were
binned by a factor of two, which gives a scale of 0\farcs162/pixel.
Thus the  field of view is $\sim$ 5\farcm5 $\times$ 5\farcm5. 
During the photometric night 43 images ( 3 $\times$ 1 min,  40 $\times$ 3 min) 
resulting in an integration time of 123 min in total were taken. \p0537 was
placed close to the center of one of the two CCDs (CCD \#45, which covers the
eastern part of the mosaic). The short
individual images were necessary to avoid saturation of the BL Lac
nucleus. In order to use the frames for fringe correction and superflat
purposes, jittered images were taken. 
We used a random walk jitter pattern within a rectangular box of 
$30\arcsec$ border length centered on the central position.
During the night standard stars from Landolt (\cite{lan92})  
were observed to set the zero point. 

The individual images were first corrected for bias and then for 
pixel-to-pixel variations. For the latter  we used twilight flatfields. 
A superflat was created from the 43 bias-subtracted and 
flatfielded images to correct for large-scale gradients. To remove
the fringes a scaled fringe template derived from I-band 
images taken during this  night was subtracted. 
Finally, the images were cleaned of cosmic ray hits, aligned and summed. 
The resulting image has a FWHM of 0\farcs85. 

Additionally, a short R-band image of \p0537 has been taken with the VLT-UT1
and FORS on the night October, 10-11 1999 for the preparation of 
the spectroscopic observations. The data reduction was similar as for the
I-band images.

\subsection{Spectroscopy}
      
Low-resolution spectra of $\sim$ 46 objects in the field of \p0537 were
collected with FORS1 at the VLT1 (Antu) during two clear nights
on November 19-21, 2000. Observations were done in MOS (Multi-Object
Spectroscopy) mode, in which FORS1 provides 19 slitlets of $\sim$ 22\arcsec 
length in a vertical array. Two MOS-setups were designed. Setup A was observed
for 1 hour (3 $\times$ 20 min) at PA = $49\fdg6$ (counted counterclockwise
from N), whereas setup B 
was observed for  2 hours (3 $\times$ 40 min) at PA = $32\fdg6$. 
We used the grism 150I, which gave us a spectral scale  of $\sim$ 
5.5 ${\rm \AA}$/pixel. The slit width was set to 1\arcsec. 
The wavelength range covered was $\sim {\rm 4000 - 10000  \AA}$. 
Slitlet \#9 for setup A
covered two  brighter nearby companion objects of \p0537 
(labeled A1 and A2 in Fig. \ref{fighost}),
whereas the slitlet \#9 for setup B covered two 
fainter nearby companion galaxies of \p0537
(labeled B1 and B2 in Fig. \ref{fighost}).
The remaining slitlets were filled serendipitously with 
galaxies in the field in order to search for galaxies at the redshift 
of \p0537. In some cases we were able to put two objects onto one slitlet.
The mean DIMM seeing was 0\farcs6 FWHM for setup A and 
1\farcs05 FWHM for setup B. Finally, a short (10min) long-slit spectrum
across the nucleus of \p0537 at PA = 0 with the same slit width and grism
was observed on the night November, 21-22 during a period of poor seeing
conditions (1\farcs8 FWHM). At the end of each of the nights, 
spectrophotometric standards from Oke (\cite{oke90}) were observed.

The data reduction of the individual spectra (bias subtraction, flatfielding,
cosmic ray removal, sky subtraction, wavelength calibration, etc.) was
performed using standard MIDAS routines. The individual spectra of each 
slitlet were averaged and smoothed using a running mean 
with a width of 3 pixels. The FWHM spectral resolution measured from strong
night sky emission lines is $\sim 25 {\rm \AA}$.

\section{Analysis and results}

\subsection{Imaging}

\subsubsection{The host galaxy and nearby environment}

In order to search for the host galaxy or a galaxy along the
line-of-sight to \p0537 we applied a fully 2$-$dimensional fitting
procedure to the image (see Nilsson et al. \cite{nil99} and Heidt et
al. \cite{hei99} for details).  We first extracted a PSF by averaging
the three brightest stars in the field of \p0537. Only stars present
on the same CCD as \p0537 were used. Unfortunately, \p0537 was bright
during the observations. Thus, even the brightest stars in the field usable
for the extraction of the PSF were at
least 1.5 mag fainter than the BL Lac itself.  
In order to improve the S/N in the outer parts of the PSF, we smoothed
the PSF outside the radius of 2\farcs4 ($\sim 3$ FWHM) 
using a Gaussian kernel. In
the inner part (r $<$ 2\farcs4) no smoothing was applied in order to
accurately preserve the rapidly changing brightness profile close to
the center of the PSF.  This composite PSF was then used for the model
fits.

\begin{figure*}[t]
\centerline{\hbox{
\psfig{figure=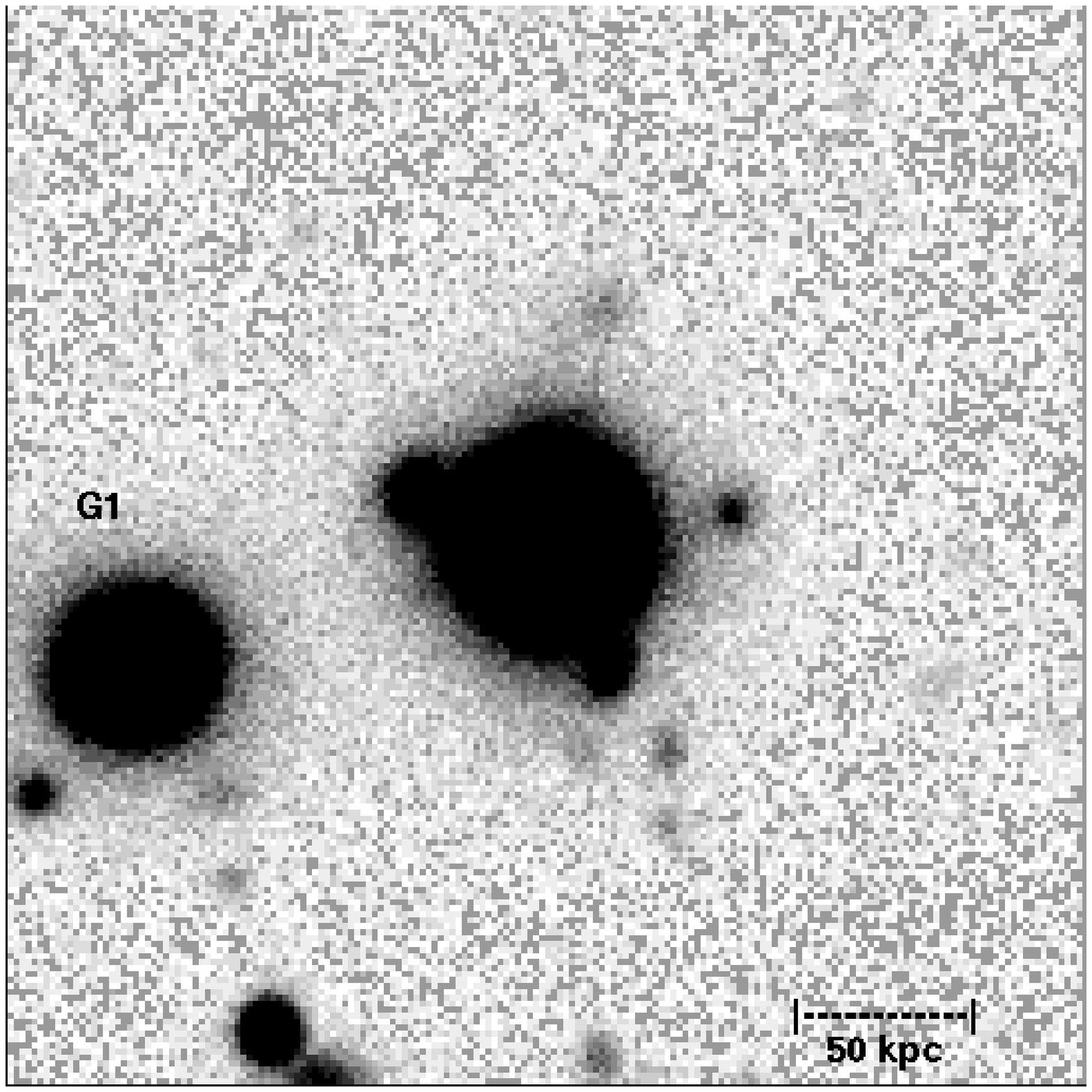,width=8cm,clip=t}
\hspace*{.2cm}
\psfig{figure=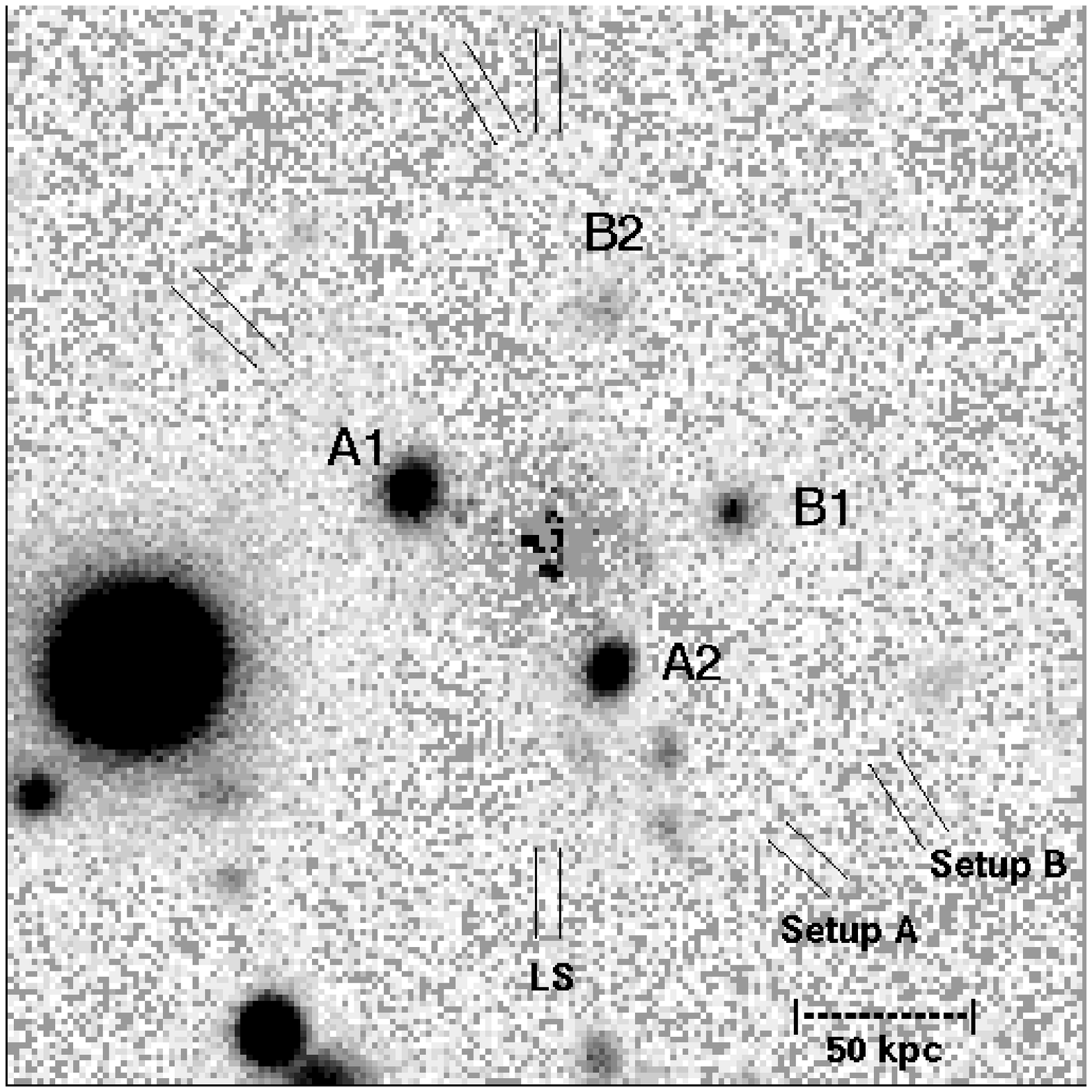,width=8cm,clip=t}
}}
\caption [] {Left) NTT+SUSI2 I-band image of \p0537. 
FOV is 30\arcsec $\times$
  30\arcsec. North is up, east to the left, FWHM $\sim 0\farcs85$. 
The galaxy G1 11\arcsec\ to the
  east of \p0537 at z = 0.186 is discussed in Stickel et al. (\cite{sti88}).
  Right) Same image after subtraction of a scaled PSF. 
The four closest companion galaxies to \p0537 are labeled.
Labeling for A1, A2 and B1 follows
Falomo et al. (\cite{fal92}) and Lewis \& Ibata (\cite{lew00}). B2 is a new
detection by us. The orientation of the MOS setups A and B as well as of the 
longslit spectrum is also indicated.}
\label{fighost}
\end{figure*}

We fitted three different models
to the BL Lac. One representing the AGN (scaled PSF), one representing
an AGN + bulge (scaled PSF + convolved de Vaucouleurs model) and one
representing an AGN + disk (scaled PSF + convolved disk model). We did
not allow for any offset between the AGN and model galaxy, the
position angle and ellipticity of the putative host galaxy were set to
zero. Prior to fitting, we masked carefully all regions affected by
  nearby companion objects. 

\begin{figure}[]
\centerline{\hbox{
\psfig{figure=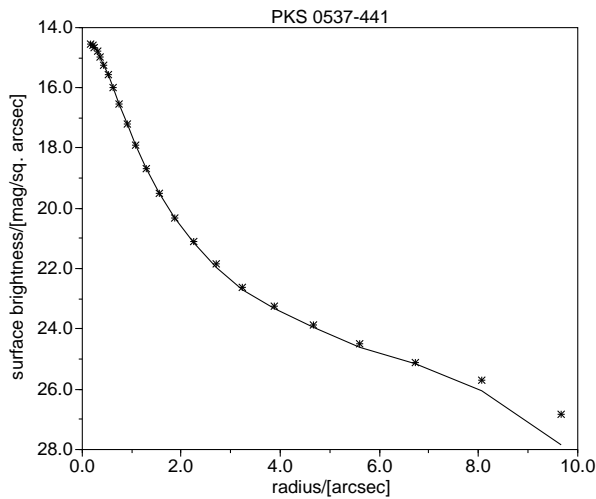,width=8cm,clip=t}
}}
\caption [] {Surface brightness profile of \p0537 (asterisks) compared
to a scaled PSF (solid line).} 
\label{fighost1dim}
\end{figure}

Already the fit with a scaled PSF was an adequate description of the observed
light distribution of \p0537, i.e. we could neither detect the host
galaxy of \p0537 nor a potential galaxy along the line of sight to the BL
Lac. \p0537 appears completely unresolved. Its brightness was \mi = 14.45,
which is relatively bright  for this source. 
The image of \p0537 and its residuals after
subtraction of the best fit AGN is shown in Fig. \ref{fighost}, the 1-dim
profiles of \p0537 and a scaled PSF in Fig. \ref{fighost1dim}. 

As can be seen from Fig. \ref{fighost}, there are residuals in the center 
of \p0537, which are typical for PSF-subtraction. 
Remarkably, our PSF traces the surface brightness of \p0537 out to a radius of 
about 8\arcsec\ covering $\sim$12 magnitudes in brightness 
(Fig. \ref{fighost1dim}). There are some slight differences at about
2-3\arcsec\ from the core (seen also as
remaining light to the north and west
of the BL Lac in Fig. \ref{fighost}), whereas at radii $>$ 8\arcsec\ the 
PSF clearly underestimates the flux from \p0537. Both features are due to 
to the imperfect PSF.

We have estimated upper limits for the host galaxy brightness
using simulated images of \p0537. The simulated images consisted
of a core component with \mi = 14.45 and a host galaxy with effective
radius r$_{\rm eff}$ = $0\farcs95$ (10 kpc at z = 0.893). Using
progressively fainter host galaxies we determined the highest host
magnitude that allowed us to make a host detection. For the host
galaxy to be detected we required r$_{\rm eff}^{av}$ $>$ $5\sigma_{\rm
reff}$, where r$_{\rm eff}^{av}$ is the average effective radius and
$\sigma_{\rm reff}$ is the rms scatter in a set of 30 simulations.
Using this method we estimate that a host galaxy with \mi $<$ 19.0
would have been detected.

Inspection of Fig. \ref{fighost} reveals four resolved galaxies
within 7\arcsec of \p0537. The three brightest ones
(A1, A2 and B1)
have already been detected by Falomo et al. (\cite{fal92}) and Lewis \& Ibata 
(\cite{lew00}). In addition, there is a fourth companion galaxy 
$\sim$ 6\farcs5 
to the north of \p0537 (labeled B2 in Fig. \ref{fighost}), which is also
barely visible in Fig. 1 of Falomo et al. (\cite{fal92}). The  brightness
of these objects is \mi $\sim$ 21.0, 21.2, 22.0 and 23.0 and 
their projected distances from the BL Lac are
3\farcs9, 3\farcs8, 5\arcsec and 6\farcs4 for A1, A2, B1 and B2, respectively.

\subsubsection{Cluster environment}
\label{cluster}

\begin{figure*}[t]
\centerline{\hbox{
\psfig{figure=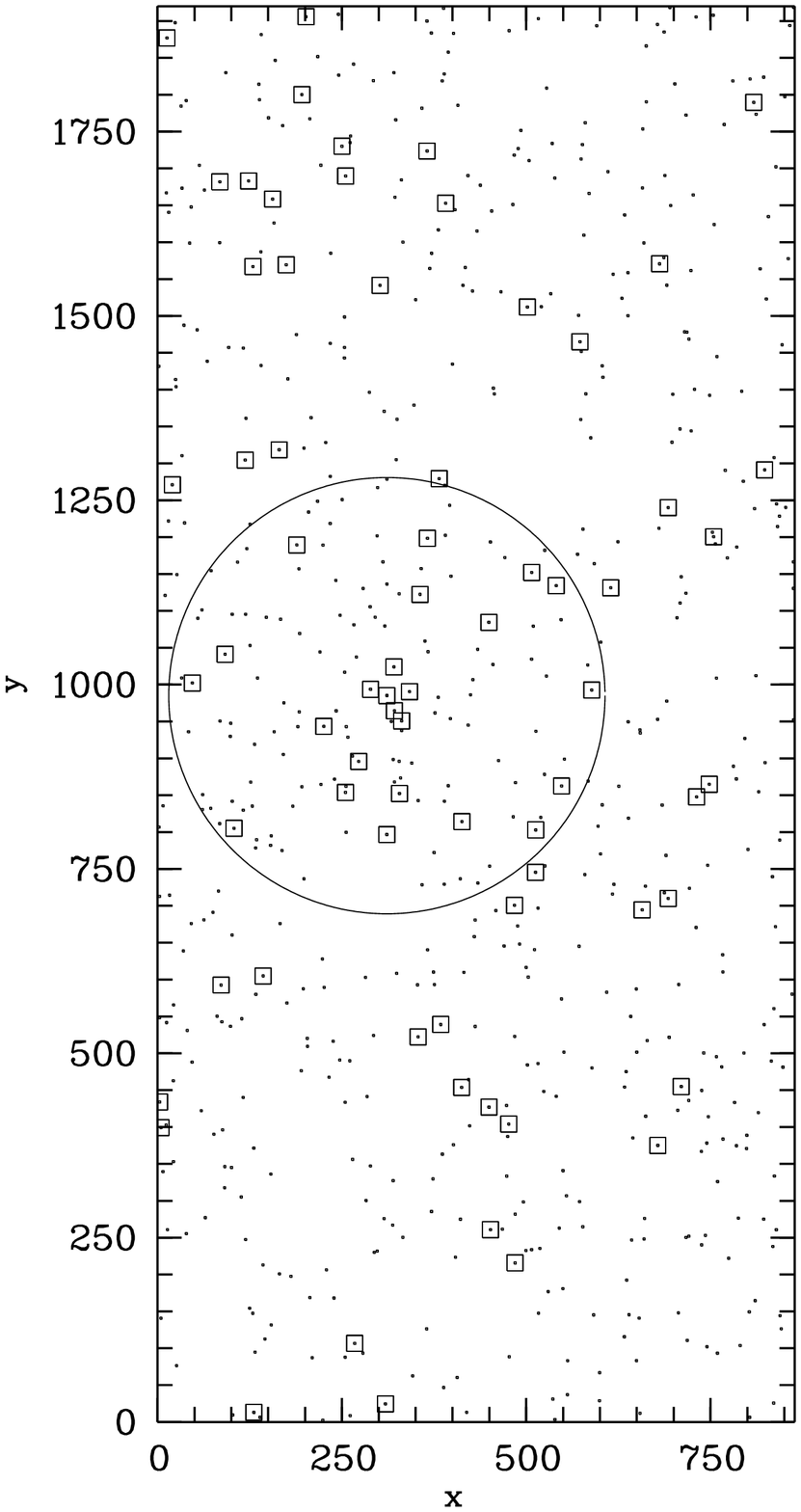,width=8cm,clip=t}
\hspace*{.2cm}
\psfig{figure=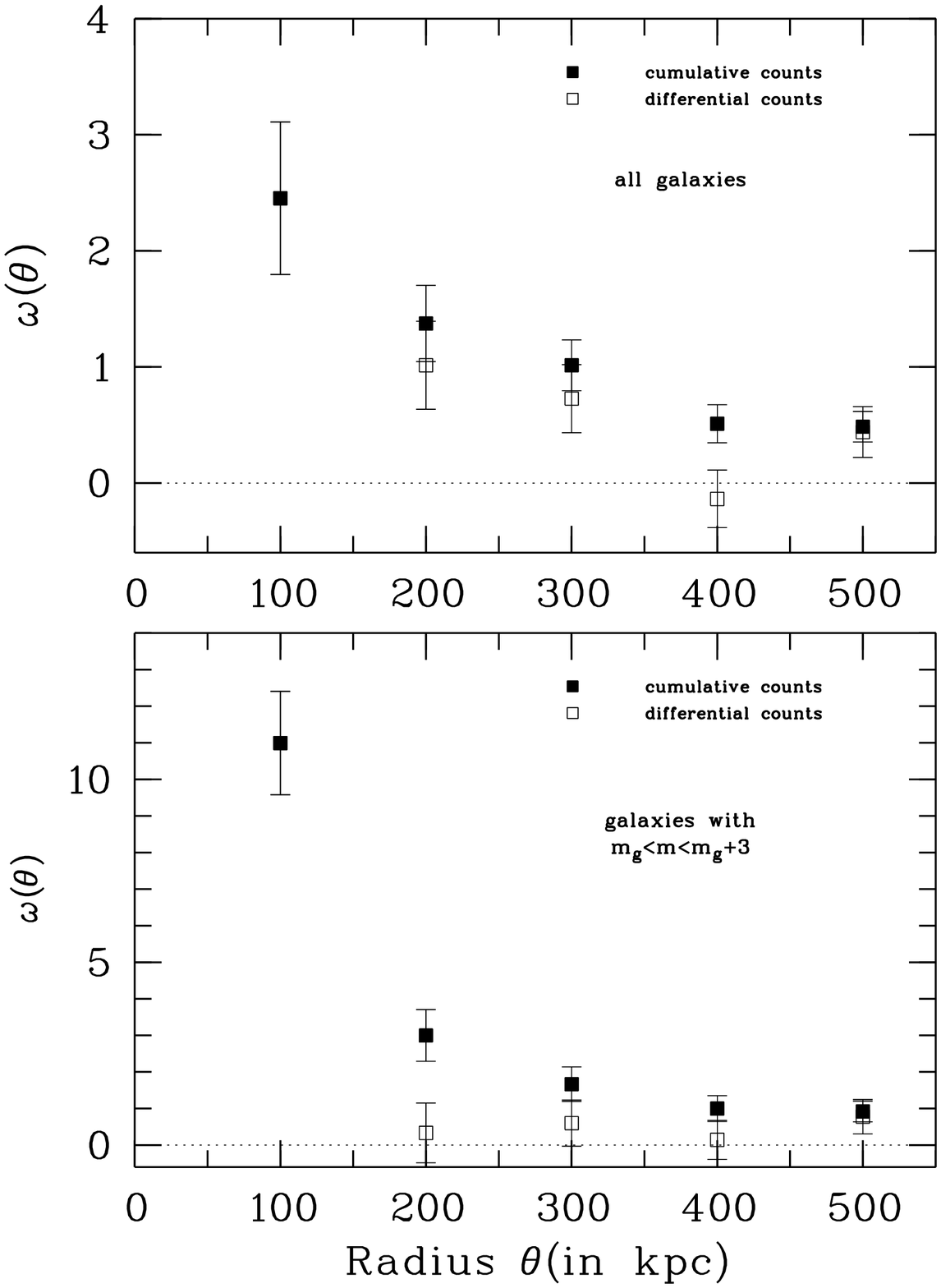,width=8cm,clip=t}
}}
\caption [] {Left) The field around \p0537 from the NTT I-band
  image. 
The 500 kpc environment is indicated by a circle centered on \p0537.
Plotted are all objects with CLASS\_STAR$<$0.5 (i.e.~galaxies).  
Galaxies in the magnitude range $m_{g} < m < m_{g}+3$ are 
marked by a square. \p0537 is included as a galaxy, 
according to the magnitude-redshift
relation by Eales (\cite{eales}) we used ${\rm m}_{\rm g}({\rm I})$ = 19.5. 
A density excess towards the BL Lac can easily
be  seen. There are 24 galaxies within and 47 galaxies 
outside the 500kpc region, i.e. 1/3 of the galaxies are within 1/6 of
the area covered by this CCD.
Right) The galaxy density excess $\omega (\theta)$ 
as a function of radial distance to the BL Lac. 
The mean field galaxy density is indicated by the horizontal line. 
It has been
derived from the entire field (i.e. both 2k$\times$4k CCDs) excluding
the 500kpc environment of the BL Lac. The top
panel shows  $\omega (\theta)$ for all galaxies, whereas in the
bottom panel only galaxies
within the range $m_{g} < m < m_{g}+3$ have been used. 
For both diagrams only galaxies (CLASS\_STAR$<$0.5) are considered.}
\label{figcluster}
\end{figure*}

In order to analyze the cluster environment of \p0537 we
first created an object catalog with positions, 
magnitudes (MAG\_BEST) and a star/galaxy 
classification index (CLASS\_STAR) using SExtractor 
(Bertin \& Arnouts \cite{ber96}). MAG\_BEST is an isophotal magnitude
which considers deblending of objects and applies an aperture flux correction 
at the edges of sources. The CLASS\_STAR parameter can be in the range 
between 0 (galaxy) and 1 (point source). 
This value is calculated by a comparison of the FWHM of an image and the 
morphology of objects. It considers both the ellipticity of objects as well as 
their ratio of maximum brightness to their extension. 

Within the field of 2 (CCDs) $\times$ 2\farcm32 $\times$ 5\farcm15 
we detected $\sim$ 1200 sources. 
After a careful visual inspection of the detections, $\sim$ 60 obviously wrong
entries were removed (false detections close to the edges of the frame
or imperfectly removed cosmic ray hits) and a few sources were added
(which were not detected by SExtractor but clearly present). In the latter
case we performed aperture photometry with standard Midas routines.
The completeness limit derived from the peak in the 
differential galaxy number counts is \mi $\sim$ 24.5.
Finally, the projected distance (in kpc) of each source to the BL Lac 
was calculated.

To determine the strength of galaxy clustering around  
\p0537 we measured the surface galaxy density as a 
function of projected distance from the BL Lac. All sources within 
five circular annuli with equidistant radii from 100 kpc to 500 kpc 
centered on \p0537 were
counted and their resulting counts compared to those derived for 
field sources. Precisely, we calculated 
\begin{displaymath}
\omega(\theta +\Delta \theta) =
\frac{N_{B}(\theta+\Delta \theta)}{N_{F}(\theta+\Delta \theta)} -1\\\\
\end{displaymath}
where $N_{B}(\theta+\Delta \theta)$ are the number counts
within a given area $\theta+\Delta \theta$ from the BL Lac and
$N_{F}(\theta+\Delta \theta)$ are the counts which would be expected
if the galaxies are distributed as in the field.
The mean field galaxy density $N/arcmin^2$ 
was calculated from galaxy counts at $\theta > 500$ kpc 
from which $N_{F}(\theta+\Delta \theta)$ has been scaled according to
the area covered.
Both, cumulative counts as well as differential counts  
have been measured.

Fig. \ref{figcluster} (top right) shows $\omega(\theta +\Delta \theta)$ 
by using all galaxies detected (CLASS\_STAR $<$ 0.5).
Even without any restriction of the galaxies to a certain magnitude range 
a density excess towards \p0537 can be seen. 
The clustering signal becomes even stronger if we use only
galaxies in a magnitude range, which has the highest probability to 
be at the BL Lac redshift (Fig. \ref{figcluster} bottom right). 
For the definition of the magnitude range we followed Hill and Lilly
(\cite{hill}, hereafter HL91) and considered only galaxies 
within $m_{g}<m<m_{g}+3$. The typical apparent magnitude $m_{g}$ of a 
giant elliptical galaxy at a given redshift was obtained 
from the magnitude--redshift relation for radio galaxies 
of Eales (\cite{eales})
\begin{displaymath}
m_{g}(R) = 21.05+5.3 log z. 
\end{displaymath}
Since we observed in the I-band, $m_{g}(I)$ was estimated by subtracting 
the $R-I$ color of an early type galaxy (k-correction and galactic 
extinction towards \p0537 had also been taken into account).

For an estimate of the cluster richness and a direct comparison to other 
studies, we calculated the excess of galaxy counts 
$N_{0.5}$ as defined by HL91. 
This is the number of galaxies ranging from $m_{g}<m<m_{g}+3$ 
within the 0.5 Mpc radius around the BL Lac minus the average number 
counts of the field galaxy distribution.
For \p0537 we find $N_{0.5}=11.5\pm 3.5$. 
This corresponds to a spatial cross-correlation amplitude of the order of 
$B_{qg}\approx 435$ following the relation 
$B_{qg} = (37.8\pm 10.9)N_{0.5}$ given by Wold et al.~(\cite{wold}),
and hence to a cluster of Abell 
richness class 0-1 (if compared to the calibration in Tab. 4 and Fig. 3 
of HL91).

\begin{table*}[]
\caption[]{Classified objects from the 2 MOS-setups in the field of \p0537.}
\begin{center}
\begin{tabular}{l|rrccccccl}
\hline
 & & & & & & &\\
Slitlet\footnotemark[1] & $\Delta \alpha$ & $\Delta \delta$ 
& \mr & \Mr & \mi & \Mi & z & SED/AGN & comments\\
 & [\arcsec] & [\arcsec] & [mag] & [mag] & [mag]& [mag]& \\
& & & & & & &\\
\hline
& & & & & & &\\
\#a2  & 180&  22&23.7&-21.9&22.4&-23.1&1.263&3& \o2 \\
\#a3  & 158&  13&23.5&-20.2&22.4&-21.3&0.658&2& \o2\\
\#a4a &  93&  65&18.7&-21.6&17.9&-22.3&0.166&1& Ca K+H, g-band, Mg b, Na D  \\
\#a4b &  88&  61&23.3&-20.2&22.4&-21.0&0.604&3& \o2\\
\#a6  &  63&  24&22.9&-17.4&    &     &0.165&5& \o2, \O3\\
\#a7  & -66& -13&23.6&-19.1&    &     &0.443&5& \o2, \hb, \O3\\
\#a8  & -11&  48&22.4&-19.9&21.7&-20.6&0.387&4&  \o2, \hb, \O3\\
\#a9a &  4 &   1&    &     &21.0&-23.5&0.885& QSO& Mg II, BAL QSO?, A1  \\
\#a9b &   0&   0&    &     &14.5&-30.0&0.893& BL Lac& Mg II, \o2, \p0537  \\
\#a9c &  -2&  -3&    &     &21.2&-23.5&0.947&2& Ca K+H, Mg b, A2  \\
\#a10 &   0& -31&23.4&-21.1&22.2&-22.3&0.892&3& \o2  \\
\#a11 & -39& -20&23.0&-15.3&22.5&-15.8&0.068&3& \hb, \O3, \ha\\ 
\#a12 & -12&-100&23.9&     &22.1&     &     & & M-star \\
\#a13 & -24&-117&23.3&-21.4&22.1&-22.6&0.952&3& \o2\\  
\#a15a&-146& -47&23.5&-19.6&    &     &0.529&3& \o2\\
\#a15b&-153& -53&20.6&-22.6&    &     &0.555&3& \o2, \O3\\
\#a17 &-136&-130&23.7&-21.5&    &     &1.115&4& \o2\\
\#a18 &-165&-130&24.0&-21.0&    &     &1.058&3& \o2\\
& & & & & & &\\
\hline
& & & & & & &\\
\#b2  & 141&  97&24.3&-20.6&23.0&-21.8&0.998&3& \o2\\
\#b3a & 132&  84&23.5&-19.8&22.6&-20.6&0.560&3& \o2\\
\#b4  &  19& 131&23.3&-21.4&22.1&-22.6&0.936&2& \o2, Ca K+H\\
\#b5  &  23&  94&23.7&-20.9&22.1&-22.5&0.916&2& \o2, Ca K+H, g-band \\
\#b6  &  66&  43&23.7&-18.9&    &     &0.424&5& \o2, \hb, \O3\\
\#b7  &  46&  26&24.3&-19.2&23.2&-20.2&0.604&5& \o2, \O3  \\
\#b8  &  31&  17&24.3&-19.4&23.2&-20.5&0.652&5&  \o2, \hb, \O3\\
\#b9a &  -5&   1&    &     &23.0&-21.5&0.892&5& \o2, B2 \\
\#b9b &  -1&   6&    &     &22.0&-22.5&0.895&2& \o2, Ca K+H, g-band, B1 \\
\#b10 &  -6& -15&24.5&-19.3&23.7&-20.0&0.672&5&  \o2, \O3 \\
\#b12 & -34& -46&24.9&-19.8&23.5&-21.1&0.927&4& \o2 \\
\#b13a& -19& -84&24.3&-20.4&22.8&-21.9&0.949&3& \o2 \\
\#b14a& -65& -85&22.1&-21.2&21.3&-22.0&0.564&3& \o2, \O3 \\
\#b14b& -67& -88&23.8&-20.7&23.3&-21.2&0 892&4& \o2 \\
\#b16 & -74&-144&23.5&-21.2&22.4&-22.3&0.948&1& Ca K+H, g-band\\
\#b17a& -85&-162&23.5&-21.4&    &     &0.992&1& Ca K+H, g-band\\
\#b17b& -86&-164&23.0&-21.8&    &     &0.989&1& Ca K+H, g-band\\
\#b18 & -90&-173&21.0&     &    &     &     & & M-star \\
& & & & & & &\\
\hline
\end{tabular}
\end{center}
\vspace*{.2cm}
\begin{minipage}[]{18cm}
{$^1$Integration time was 1 hour for the \#a$\ast$ objects and 2 hours for the 
\#b$\ast$ objects. In some cases, two objects could be placed onto the 
slitlet. Those objects were labeled $\ast$a and $\ast$b, respectively. 
}
\end{minipage}
\label{ztab}
\end{table*}

\subsection{Spectroscopy}

Preliminary results of the spectroscopic observations have been
described by us in Heidt et al. (\cite{hei2001a}, \cite{hei2001b},
\cite{hei2003}). 
These were solely based on the identification of 
emission/absorption lines in the
spectra. For the present paper more accurate redshifts could be 
determined by a more sophisticated analysis, which further allowed us to 
roughly classify the objects into galaxy types. 

\begin{figure*}[]
\centerline{\hbox{
\psfig{figure=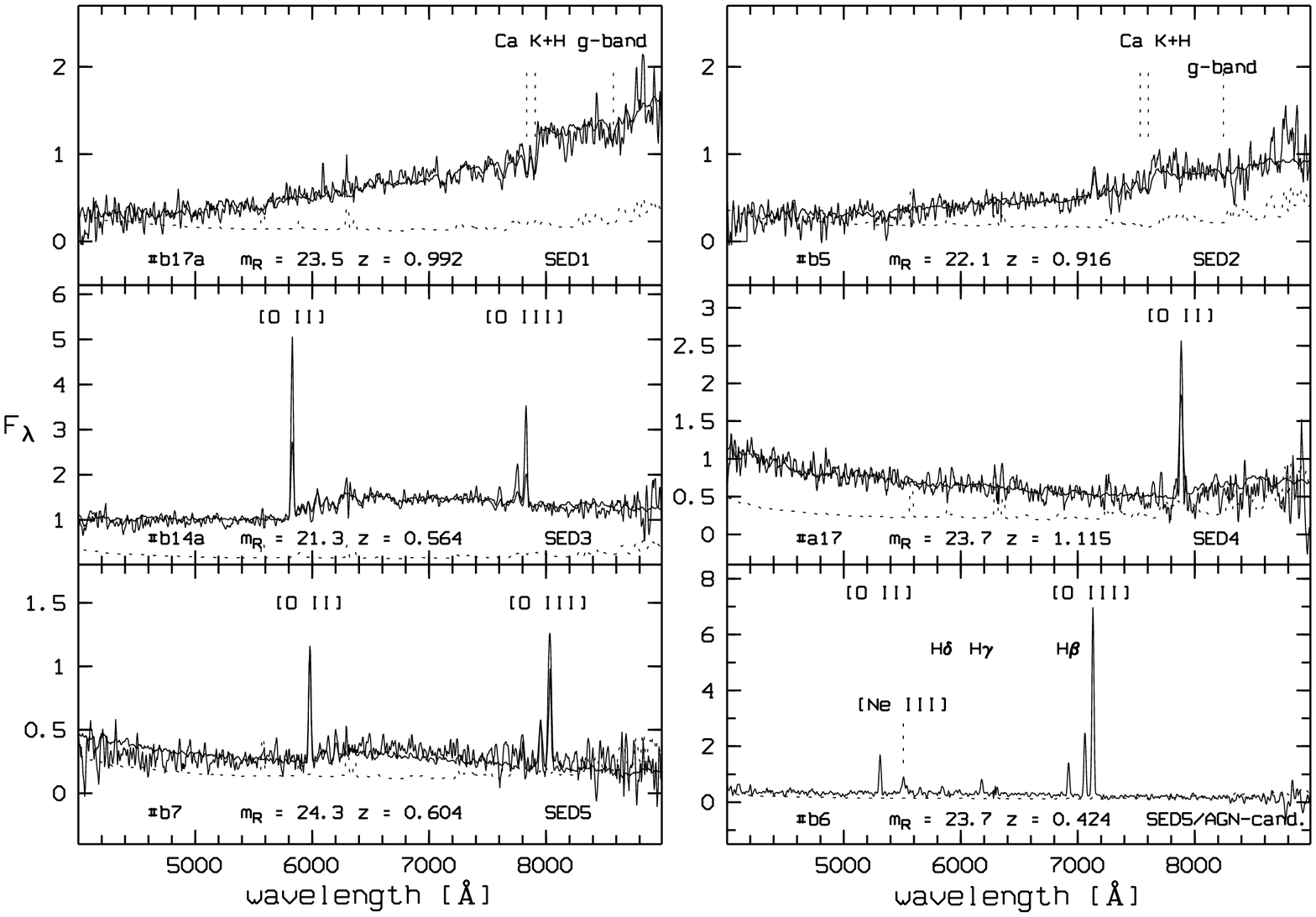,width=18cm,angle=-90,clip=t}
}}
\caption [] {
Spectra of representative galaxies and of an AGN candidate detected in the
field of \p0537.  Templates of average spectral energy distributions of
galaxies of different Hubble type (SED1-5) are overplotted on the observed
galaxies' spectra. 
The error spectrum (dashed line) is displayed as well.}
\label{figspec}
\end{figure*}

The spectra were first corrected for the atmospheric A-band and B-band 
using the standard stars during the same nights. This is
very important, since the A-band coincides at z $\sim$ 0.9 (the BL
Lacs redshift) with the 4000 \AA\ break in early-type galaxies. Then we
searched for obvious emission/absorption line features in our
spectra and determined their centroids by fitting Gaussians to them.
In order to avoid misidentifications due to artefacts 
(e.g. imperfect removed cosmic ray signatures), 
all the 20 min and 40 min spectra were inspected and compared individually.
Finally, we fitted interactively five different spectral energy 
distributions (SEDs) to the spectra. 
As starting values we used preliminary 
redshifts obtained from emission/absorption lines.
The five different SEDs used for the fitting procedure 
were derived from spectroscopic
observations of about 500 galaxies in the FORS Deep Field\footnote{see Heidt
  et al. (\cite{heifdf}) for details on the FORS Deep Field} (S. Noll, in
prep.). They have the advantage that they were derived with the same
spectroscopic setup, instrument and telescope as the 
observations presented here. 
The galaxy types covered are: SED1 =
E-S0, SED2 = Sa-Sc, SED3 = Sd/Irr, SED4 = starburst, SED5 = extreme
starburst. Four of the SED5 sources (\#a6, \#a7, \#b6 and \#b10)
may be AGN. They all have the commonly used flux ratios \O3 5007/\o2 $>$ 1, 
\O3 4959/\hb\ $>$ 1 and \O3 5007/\hb\ $>$ 3 for a classification as an AGN. 
Unfortunately, due to the 
relatively high redshift of the sources, the important diagnostic lines 
$[$N II$]$ 6583/\ha\ are either in the extreme red part
of the spectrum dominated by night sky emission or outside the spectral range
covered. The ratio $[$O III$]$ 5007/\hb\ vs. $[$ N II$]$ 6583/\ha\ is a 
robust diagnostic diagram to separate starburst
galaxies from AGN (Veilleux \& Osterbrock \cite{oster}). 

The approach of fitting SEDs is especially helpful for the
redshift determination for
galaxies dominated by an old stellar population. The error in the
redshift determination is $\Delta$z $\sim$ 0.001 for sources with at least 2 
emission lines and $\Delta$z $\leq$ 0.003 for the remaining sources. The
latter error was derived by slightly varying the redshift of the
determined SEDs for individual galaxies until an obvious mismatch was
apparent. 

\begin{figure}[]
\centerline{\hbox{
\psfig{figure=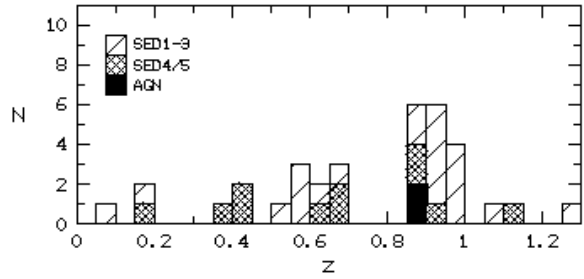,width=8cm,clip=t}
}}
\caption [] {Distribution of the objects from the 2 MOS-setups
with redshift and type. Hatched = SED1-3, double hatched = SED4/5 and black = 
AGN, respectively.}
\label{fign_z}
\end{figure}

The results are summarized in Tab. \ref{ztab}. Column 1 
identifies the slitlet, where the objects were located, columns 2 and
3 give the offsets with respect to \p0537 followed by the apparent and
absolute R and I magnitudes in columns 4 to 7. 
The apparent R and I magnitudes were measured using SExtractor 
(Bertin \& Arnouts \cite{ber96}). 
Since \p0537 is strongly saturated on the 
R-band acquisition image taken for the preparation of the MOS setups,
we do not give R magnitudes for the companion
galaxies A1, A2, B1 and B2. The R- and I-images do not fully
overlap. Hence, I magnitudes could not be derived for all sources.
No k-correction was applied to the absolute magnitudes, because the
classification into SEDs is very rough.
Column 8 gives the redshift and column 9 the rough SED classification.
Finally, in column 10, the detected lines are listed and comments are given.
In Fig. \ref{figspec} representative spectra of objects with SED1-5 and one
AGN candidate are shown.

For 36 out of the 46 spectra in total, we were able to determine the nature 
of the sources. Their R and I magnitudes range from 18 to 24.5, 
most have \mr $>$ 22.5. Their absolute magnitudes range from 
-15.3 to -22.6 (median -20.6) in R and from -16 to -30 (median -21.9) in I.
Remarkable is the extreme brightness of the BL Lac (-30.0) and brightness
of the two nearby
companions A1 (-23.5, QSO) and A2 (-23.5, early-type at z = 0.947).
We found 32 galaxies (4 SED1, 5 SED2, 12
SED3, 4 SED4 and 7 SED5), 2 AGN (\p0537 and A1, which will be discussed in the
next section) and 2 stars. 

For about 60\% (19/34) of the galaxies, at least two emission/absorption
lines could be detected, which allowed a secure redshift measurement. 
In 14 sources we could only detect a single narrow emission line, 
which we identified with \o2 in all cases. 
This identification was supported in most cases by either 
a jump of the continuum redwards of the \o2 line indicating the 
4000 ${\rm \AA}$ break or a blue continuum typical for strong star 
forming galaxies. Finally, in companion galaxy
A1, we detected a broad emission line which we identified as Mg II.

\begin{figure*}[]
\centerline{\hbox{
\psfig{figure=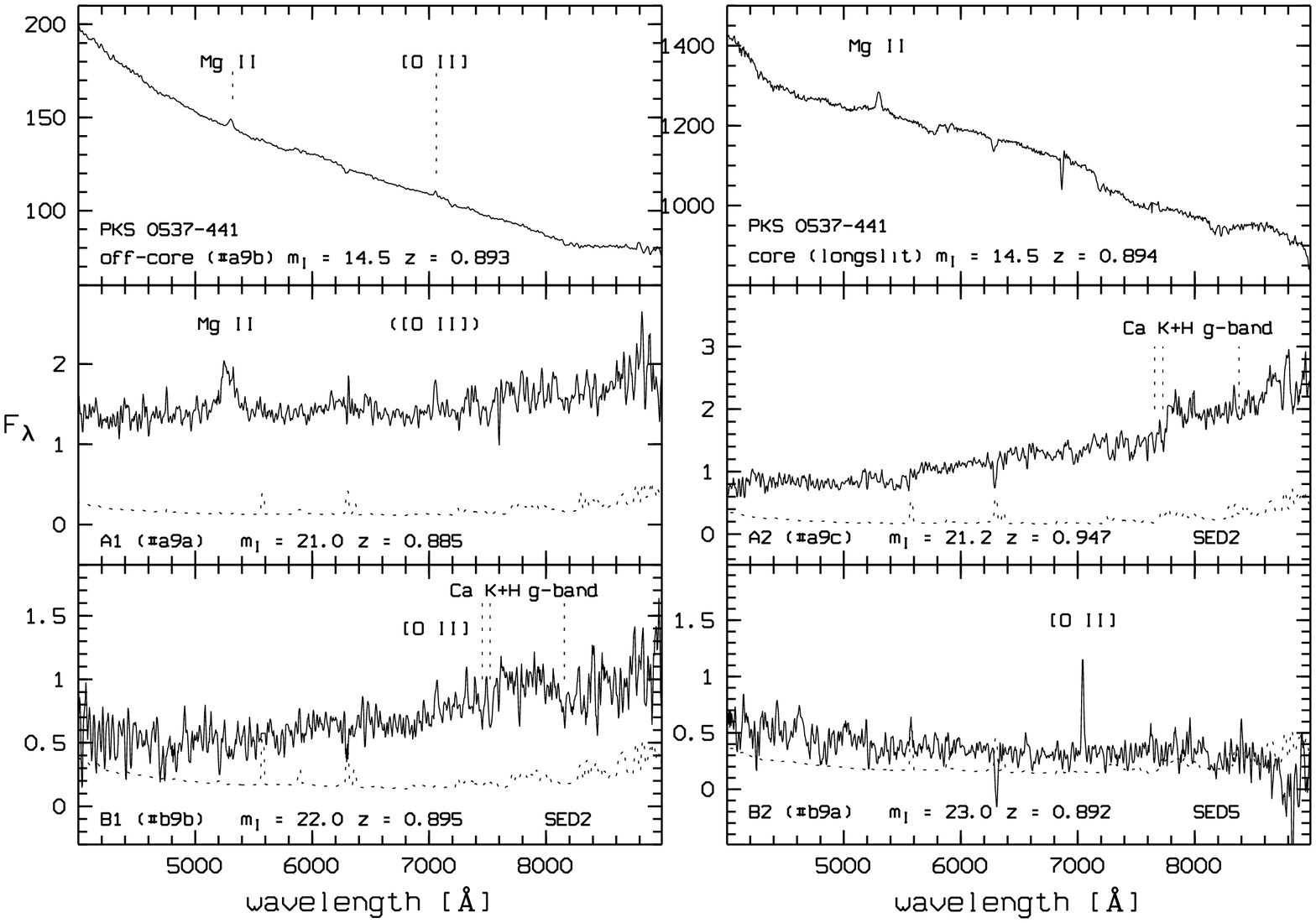,width=18cm,angle=-90,clip=t}
}}
\caption [] {Spectra of \p0537 (off-nucleus from setup A and across the core
  from longslit observations) and of the 4 companion galaxies A1, A2, B1 and
  B2. For the companion galaxies the error spectrum (dashed line) is included.}
\label{figspeccomp}
\end{figure*}

The  redshifts of the objects range from 0.07 up to 1.3, 
19 are at z $>$ 0.8. Most objects are dominated by \o2 emission, 
implying relatively high star formation rates. We found also 
several early-type galaxies at z $>$ 0.9.
Four objects have redshifts similar to \p0537 ($\Delta$ z $<$ 0.002). 
The distribution of objects with redshift and type is
shown in Fig. \ref{fign_z}.

\subsubsection{\p0537 and  A1, A2, B1 \& B2}
\label{compspec}

In Fig. \ref{figspeccomp} the 1-dim spectra of \p0537 and the 
companion galaxies A1, A2, B1 and B2 are
displayed. Although the companion galaxies are about 6-8 mag fainter 
than the BL Lac, their spectra were sufficiently well separated from \p0537 
to measure not only characteristic lines, but also to detect the continuum. 
The spectra are discussed below.

\p0537: In the off-nuclear spectrum of \p0537,  Mg II
at 5300 ${\rm \AA}$ and \o2 at 7052 ${\rm \AA}$ are detected, which give z =
0.893 on average. In the spectrum across the nucleus,
only Mg II at 5300 ${\rm \AA}$ (z = 0.894), but not \o2 
was unambiguously detected. Both lines have previously been found by Lewis 
\& Ibata (\cite{lew00}) on their nuclear spectra with higher resolution. 
They give z = 0.892 in good agreement with our measurement. We also note that
\o2 has not been detected by Stickel et al. (\cite{sti93})
and by Falomo et al. (\cite{fal94}), although it may be
present in the spectrum shown by Stickel et al. (cf Fig. 2 in their paper). 

\begin{figure*}[]
\centerline{\hbox{
\psfig{figure=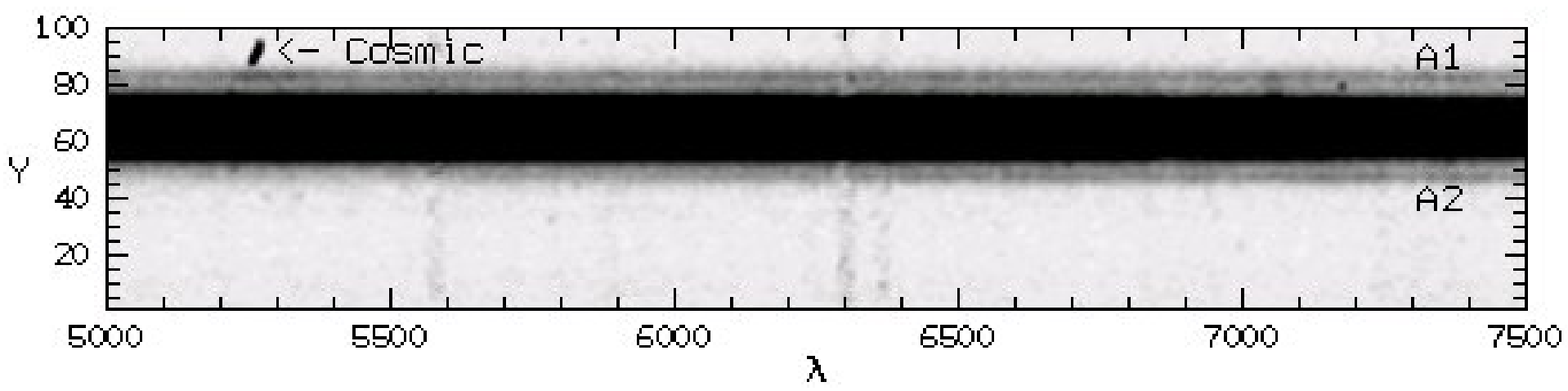,width=18cm,clip=t}
}}
\centerline{\hbox{
\psfig{figure=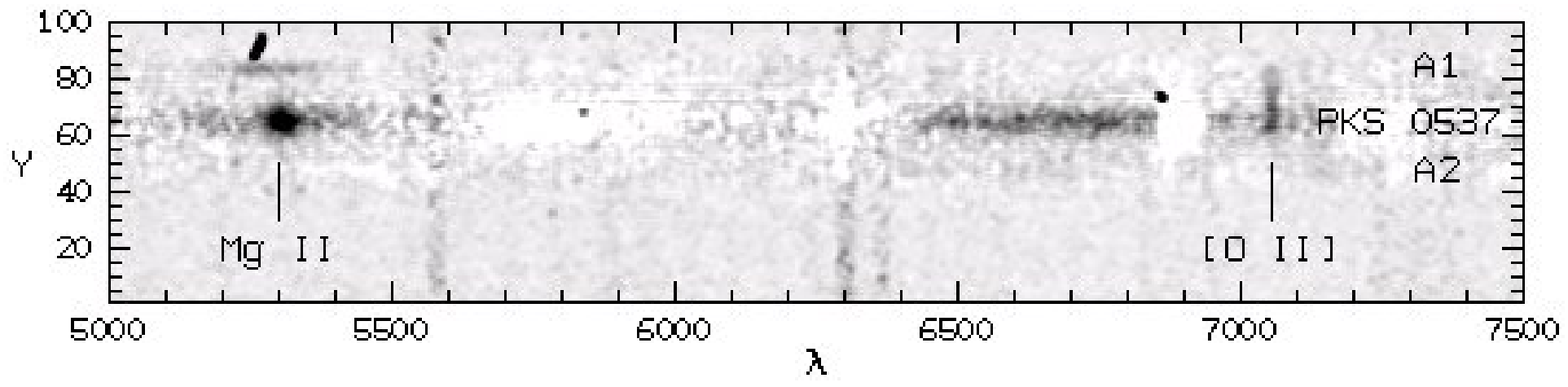,width=18cm,clip=t}
}}
\caption [] {Top) 2-dim spectrum of slitlet \#9 from setup A. The
   positions of the companion galaxies A1 and A2 are displayed. 
  A cosmic ray hit (C.) is also
  indicated. The spectra of both companion galaxies are well separated from
  the off-nuclear component of \p0537. Note the relatively blue continuum of
  A1 and the relatively red continuum of A2. The broad Mg II emission at $\sim$
  5275 ${\rm \AA}$ in A1 and some faint extended \o2 emission at 
7052 ${\rm \AA}$
  from the BL Lac towards A1 can be seen. Bottom) The same spectrum after
  subtraction of the continuum. Note that the Mg II emission from A1 
  is well separated from  the Mg II emission of \p0537, whereas the \o2 
  emission
  extends from the BL Lac towards A1. Note also that the \o2 emission 
does not extend towards the opposite direction. See text for details.}
\label{fig2dim}
\end{figure*}

A1: The blue spectrum of this object shows a very
broad asymmetric line at 5275 ${\rm \AA}$ and a narrow line 
at 7052 ${\rm \AA}$, which are most likely Mg II and \o2. 
The Mg II line is $\sim$ 25 ${\rm \AA}$ blueshifted with respect
to the Mg II line in \p0537 and gives z = 0.885. 
It can not be Ly$\alpha$ at z = 3.338 since the continuum extends
towards the blue and no obvious Ly-break is present. 
We emphasize here that the broad asymmetric line at 5275 ${\rm \AA}$ can 
also not be scattered light from the \mg2 line in \p0537 due to 
reflections inside FORS. This line in A1 can always be seen with a 
similar structure in all 3 individual images. Moreover, 
reflections inside FORS normally scale as $10^{-4}$
(W. Seifert, priv. com.), whereas we observe here a difference of the \mg2
lines of the order of $10^{-1}$.
However, the center of the \o2 line in A1 is at the 
same position as the \o2 line in the off-nuclear spectrum of \p0537. 

In order to investigate the apparent discrepancy between the Mg II and \o2
lines, we subtracted the continuum on the 2-dim spectrum of A1, A2 and
\p0537. This was obtained by fitting and subtracting row-by-row 
a smoothed cubic spline from the 2-dim spectrum. Great care was taken to match
the continuum next to 5300 ${\rm \AA}$ and 7050 ${\rm \AA}$ as well
as possible. In Fig. \ref{fig2dim} the 2-dim averaged, smoothed spectrum 
and the resulting continuum-subtracted spectrum are shown. 
There are several remarkable features. First of all, the Mg II line of
\p0537 and the Mg II line of A1 are well separated. It is also obvious, that
the Mg II line of A1 is very broad and its center somewhat blueshifted with
respect to the Mg II line in the BL Lac. By contrast, the \o2 line of \p0537
extends from the central position (at about 1\farcs7 from the core) 
with decreasing intensity without any 
blueshift towards A1. The projected length of that feature is about 4\arcsec
($\sim$ 40 kpc at z = 0.893). There might also be some (kinematic) 
substructure (wiggles) present, but this could  be due to the 
continuum-subtraction procedure. Since the central position of Mg II and \o2
in the BL Lac give the same redshift within the errors contrary to the central
position of Mg II in A1 and since the intensity of the \o2 line decreases
towards A1 without any blueshift, we believe that \o2 is intrinsic to \p0537.
Curiously, \o2 does not extend in the opposite direction. 
A physically plausible explanation for this complex configuration 
will be discussed in the next section.

A2: The spectrum of this galaxy 
is dominated by a red continuum showing Ca K+H
and Mg b absorption lines characteristic for early-type galaxies with SED2. 
No emission lines are detected. The redshift of this galaxy is z = 0.947.

B1: This galaxy has also a relatively red spectrum (SED2) showing Ca K+H and
the g-band in absorption as well as a faint \o2 emission line. The resulting
redshift is z = 0.895, close to the redshift of \p0537. 

B2: Similarly to A1, the spectrum of this source has a blue continuum (SED5)
and a relatively strong narrow emission line, which we identify with \o2.
The resulting redshift is z = 0.892, again, close to the redshift of \p0537.

\section{Discussion}

Although our I-band image of \p0537 is probably the deepest ever taken of that
field, we were not able to resolve the host galaxy of the BL Lac. This might
be at least in part due to the absence of sufficiently bright stars in the
field which  prevented us to extract a very good PSF. 
The lower limit to the host galaxy magnitude 
based on our simulations is \mi = 19.0,
which would result in \Mi = -26.8 (k-correction included). 
With a typical color of ellipticals of R-I = 0.7 (Fukugita et al. 
\cite{fukugita}) this would give \Mr = -26.1. Thus, using a typical BL Lac
host magnitude of \Mr = -23.5, our observations should have been at least 3
mag deeper given the less than optimal PSF. On the other hand, we can
rule out an intervening galaxy as bright as the early-type 
companion galaxy G1 (\mi $\sim$ 17.4) 11" to the 
east, which was seen by Stickel et al. (\cite{sti88}). This confirms
the observations by Falomo et al. (\cite{fal92}), Kotilainen et
al. (\cite{kot98}) and Scarpa et al. (\cite{scar00}). Clearly, we can not rule
out an even fainter galaxy along the line of sight (our detection limit is
about ${\rm L}^{\ast}$ at z $\sim$ 0.2). Nevertheless, there is 
no strong evidence that the properties of this very luminous and strongly
variable BL Lac nucleus
 are affected by gravitational microlensing effects. This
is further supported by the absence of absorption lines in HST FOS spectra
(Bechtold et al. \cite{bech}).
A deep high-resolution image of \p0537 in a very low state is required
to unambiguously detect the host galaxy or a faint galaxy along the line of
sight, but this is hard to accomplish.

To test, if some of the nearby companion galaxies are lensed background
sources, we have carried out multi-object spectroscopy in the
field of \p0537 including the four companion objects. None of the four
nearby companion galaxies was found to be at high redshift, i.e. it also does
not seem that the environment of \p0537 acts as a lensing system.

On the other hand, 
our spectra show that \p0537 and the companion galaxies A1, A2, B1
\& B2 form a very complex system. Whereas A2 is an early-type galaxy at z =
0.947, the companion galaxies B1 and B2 (z = 0.895 and 0.892, respectively)
are within 200 km/s systemic velocity of the BL Lac, i.e. they are very 
close to \p0537. Due to their proximity to \p0537 they could trigger the strong
activity in this source via gravitational interaction. 

The relation between \p0537 and A1 is very hard to interpret. Whereas we
detected \mg2 in both, the nuclear and off-nuclear spectrum  of \p0537, \o2 was
only detected in the latter. If one compares the relative strengths of \mg2 in
the two spectra (cf Fig. \ref{figspeccomp}), one would also expect 
to see \o2 in the nuclear spectrum, which is, however, not present. 
We also failed to detect \o2 in the nuclear 2-dim spectrum by subtracting 
the continuum in the same way as described for the off-nuclear spectrum. 
This points to the fact that the \o2 emission is only excited in that region, 
which is supported by the detection  of extended, asymmetric  \o2 emission with
decreasing intensity towards A1 in the 2-dim off-nuclear 
spectrum. 

There are several mechanisms, which may give rise to  the \o2
emission starting $\sim$ 20 kpc from the core and extending $\sim$ 40 kpc 
outwards: locally induced star formation due to 
gravitational interaction, photoionization by the central active nucleus or
shock-ionization via interaction between radio-emitting components
and the ambient gas (jet-cloud interaction). 

In order to induce star formation by gravitational interaction a companion
galaxy or at least a merger remnant is required. We think that this is
unlikely here. No companion galaxy is detected in that region
(A1 is at lower redshift)
and no obvious tidal features are present at the
location of the extended \o2 region. Moreover, elliptical galaxies normally
do not possess much gas at $\sim$ 3 \re (under the assumption that the 
host galaxy of \p0537 is a typical host galaxy with \re = 10 kpc). 

Whether photoionization by the active nucleus or jet-cloud
interaction is the dominant mechanism depends mainly on the size and 
extent of the radio sources with respect to the emission line regions
(see review by Tadhunter \cite{clive}).
Photoionization is normally the dominant mechanism for those AGN, 
where the radio source extends well beyond the emission line regions 
(Tadhunter \cite{clive}) although 
kinematical evidence for jet-cloud interaction has recently been found in 
exactly such sources (e.g. Sol\'{o}rzano-I\~{n}arrea et al. \cite{solor}). 

\begin{figure}[]
\centerline{\hbox{
\psfig{figure=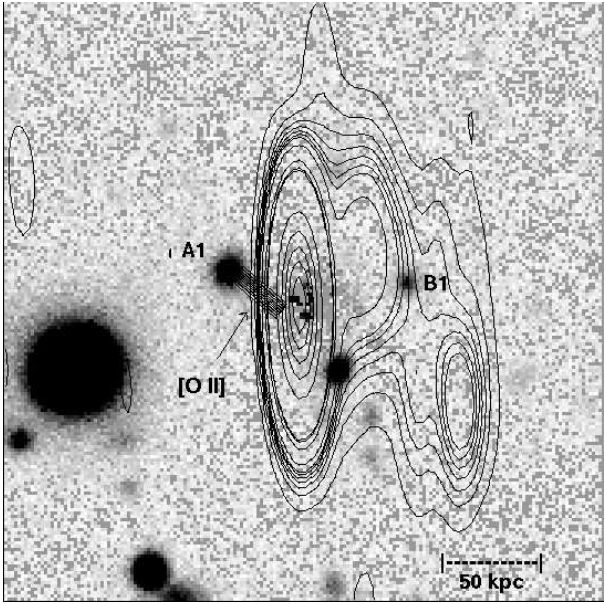,width=8cm,clip=t}
}}
\caption [] {Same as Fig. \ref{fighost} with the VLA 1360 MHz image
of \p0537 superimposed (courtesy of P. Cassaro). The region, where the
extended off-nuclear \o2 emission towards A1 has been detected is indicated. 
It is obviously on the opposite side to the jet. Note also the coincidence of
the location of B1 and the turnover of the jet.
FOV is 30\arcsec
$\times$ 30\arcsec, north is up, east to the left.}
\label{figradio}
\end{figure}

With our limited information available, we are not able to distinguish
between or set even constraints on the last two possibilities described
above. Whereas photoionization by the active nucleus is certainly 
the more realistic explanation
(extended line-emission up to 20 kpc from the nucleus has been
detected e.g. in the BL Lac PKS 0521-36, Boisson et al. \cite{boisson}),
the alternative can not be ruled out. 
\p0537 is one of the few BL Lacs classified as a FR II radio source (Rector \&
Stocke \cite{travis} and references therein) 
and shows in radio images a curved jet-like structure leading to the west 
(Cassaro et al. \cite{cassaro}). Remarkably, our 
extended \o2 emission is present almost exactly on the opposite side of
the jet of \p0537 extending $\sim$ 40 kpc in length. This is illustrated in
Fig. \ref{figradio}, where we show a comparison of the location of the jet in
\p0537 and the extended \o2 emission.
We may see here at least some signature of a
counterjet in this BL Lac. Evidence for a counterjet has been
detected via kinematic signatures of jet-cloud interaction
(line-splitting in \O3) in 3C 120 (Axon et al. \cite{axon}). 
We note, that some kinematic signatures in the extended 
\o2 emission in our 2-dim off-nuclear spectrum of \p0537 may be present,
although this may be due to the continuum-subtraction procedure. 

Further inspection of Fig. \ref{figradio} reveals that the jet in \p0537
apparently turns over close to B1. This may be due to interaction of the jet
with B1, where a pressure gradient deflects the jet considerably.
Again, whether this is correct or not can not be tested by our observations
but it is not unusual. A similar feature has been observed in the 
low-redshift BL Lac 3C 371 albeit on smaller spatial 
scales (Wrobel \& Lind \cite{lind}, Nilsson et al. \cite{3c371}).

The nature of A1 is also unclear. As we have argued in the previous
section, the emission line with some absorption towards the red is most likely
\mg2 at z = 0.885. With \Mi = -23.5 we classify A1 therefore as a QSO
(note that our I-filter samples almost exactly the B-band at z = 0.885). 
Fitting a Gaussian to the absorbed emission line gives a
width of the line of $\sim 100 {\rm \AA}$ FWHM
 ($\sim$ 5000 km/s). Corrected for redshift this gives a FWHM of $\sim 
53 {\rm \AA}$, which is is much larger as typically
measured in QSOs ($\sim 35 {\rm \AA}$ FWHM, Vanden Berk et al. \cite{berk}).
If the absorption is due to an absorbing system
 along the line of
sight to A1 one would also expect absorption in the spectrum of \p0537,
which is not observed. Therefore, the absorption must be
intrinsic to the source and A1 could hence be a  
mini low-ionization BAL QSO (LoBAL). 

\begin{figure}[]
\centerline{\hbox{
\psfig{figure=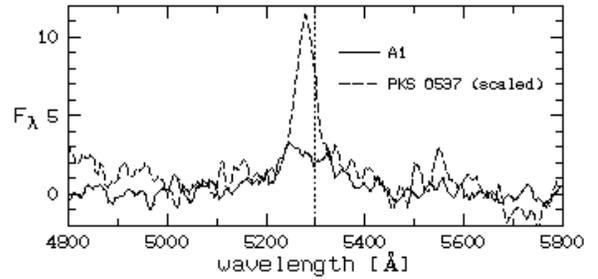,width=8cm,clip=t}
}}
\caption [] {Comparison of the \mg2 lines in A1 (solid line) and the BL 
Lac (long dashed line). The
  latter was shifted by 25 {\rm \AA} towards the blue and scaled up by a factor
  of 2. The continuum was subtracted in both spectra. Note the good match of
  the wings of the lines. For comparison the original position of the center
  of the \mg2 line in \p0537 is indicated.}
\label{figmg2}
\end{figure}

Although classical BAL QSOs have their absorption bluewards
of their systemic redshifts, a few BAL QSOs with absorption redwards
have been detected in the SDSS (Hall et al. \cite{hall}). The absorption
redwards of the systemic redshift can be either due to infalling 
gas crossing our line of sight or due to a rotation-dominated disk wind
(Hall et al. \cite{hall}). Since we do not
detect unambiguously \o2 at $\sim$ 7025 {\rm \AA} (z = 0.885) in A1 indicative
of a relatively high covering factor, infall of gas could be the dominating
absorption mechanism. There is at least indirect support for the 
identification of A1 as a mini LoBAL by the non-detection of \o2, 
which is often absent or weak
in \mg2 LoBALs (Boroson \& Meyers \cite{boroson}) or in the compact reddened
absorption-line object Hawaii 167 (Cowie et al. \cite{cowie}). 
A counterpart to A1 is
SDSS 0127+0114 (Hall et al. \cite{hall}), which is a strongly reddened
mini LoBAL and has a \mg2 configuration similar to the one observed in A1.
In Hall et al. (\cite{hall}) there is also a comparison between the dereddened
spectrum of SDSS 0127+0114 and a SDSS composite spectrum (their Fig. 12), which
shows the strength of the absorption.
We performed a similar comparison by blueshifting the
continuum-subtracted BL Lac spectrum by 25 {\rm \AA} and scaling it to match
the wings of the absorbed \mg2 emission in A1. The comparison is shown in
Fig. \ref{figmg2}. As can be seen, the spectra match very well. The
resulting FWHM would now be $\sim$ 40 {\rm \AA} instead of
$\sim$ 100 {\rm \AA}. 

We investigated the morphology of A1 on our I-band image in order to
test for the presence of an AGN. The same analysis as for \p0537 was
performed on the PSF-subtracted image (Fig. \ref{fighost}
right). According to our fits, an AGN+galaxy model is not
strongly preferred over a pure galaxy model to describe the morphology of
A1. Our best fit was an AGN+disk model with ${\rm m}_{core}$ = 22.58,
${\rm m}_{\rm disk}$ = 21.07 and 
\re = 0\farcs61 ($\chi^2$ = 1.17), whereas we got
${\rm m}_{\rm disk}$ = 20.9 and 
\re = 0\farcs45 ($\chi^2$ = 1.36) for a pure disk
model and 
${\rm m}_{de Vauc}$ = 20.59 and \re = 0\farcs64 ($\chi^2$ = 1.24) for a pure 
de Vaucouleurs model, respectively. We could not obtain a good fit with an 
AGN+de Vaucouleurs model.
On the archival HST snapshot image of \p0537 A1 and A2 are very faint, but A1 
appears more compact than A2. Thus there is no direct proof for a
compact object in the center of A1, but this may be taken as further
evidence for a (reddened) mini LoBAL QSO.

If our favored explanation for the \p0537 - A1 system is correct,
this system would form a Quasar pair. 
Based on statistical arguments Kochanek et al. (\cite{kochanek}) and
Mortlock et al. (\cite{mortlock}) have
shown that the majority of wide separation (3\arcsec $< \Delta\theta <$
10\arcsec) Quasar pairs are rather binary Quasars than gravitational
lenses. Our system with a projected separation of 3\farcs9 and with a
velocity difference of $\sim$ 870 km/s would thus fall into the category of
the rare class of binary Quasars. This is supported by the fact that
the \p0537 - A1 system is a ${\rm O}^2{\rm R}$ pair 
(both components detected in the optical, but only one in the radio; 
Kochanek et al. \cite{kochanek}), has the most extreme magnitude
difference of all QSO pairs and differs strongly in its optical
appearance. Given the relative small separation in projected distance
and relative velocity (the latter of which may even be smaller), this
system may finally merge and form a supermassive binary black hole 
(Begelman et al. \cite{begelman}).

We finally remark that the \p0537 - A1 system would now already be
the third binary QSO involving a BAL QSO (including LBQS 0103-2753
(Junkkarinen et al. \cite{junki}) and SDSS 0300+0048 (Hall et al. 
\cite{hall})). This is a high fraction of BAL QSOs amongst Quasar
binaries with respect to the relative abundance of BAL
QSOs. Curiously, all of these three Quasar binaries are at similar
redshifts (z = 0.834, 0.858, z = 0.892, 0.894, and z = 0.885, 0.893
for LBQS 0103-2753, SDSS 0300+0048 and \p0537, respectively). 
 
The analysis of the cluster environment of \p0537 
indicates that the BL Lac is located in a cluster as rich as Abell type 0-1. 
The most comprehensive study on BL Lac cluster
environments up to z = 0.65 until now
has been presented by Wurtz et al. (\cite{wur97}). 
They found that most BL Lacs reside in a relatively poor cluster environment of
Abell type $<$ 0 with an indication of evolution (richer environments) towards
higher redshifts. Similar conclusions have been drawn by Fried et al. 
(\cite{fried}) for BL Lacs from the 1 Jy sample at z $<$ 0.8. 
Thus our observations are in agreement with the observed trend.
This is also supported by the detection of four galaxies 
at very similar redshifts (within $\Delta$ z $\leq$ 0.002) as the BL
Lac. However, although the objects for the MOS-setups were selected 
serendipitously, an even higher number (10 galaxies, see Fig. \ref{fign_z}) 
was found to be in the redshift
range z = 0.9 - 1. According to the magnitude-redshift relation by Eales 
(\cite{eales}) discussed in section \ref{cluster} ${\rm m}_{\rm g}({\rm I})$
would only differ by 0.15mag between z = 0.893 and z = 0.95 (without
k-correction). We can not rule out that the enhanced galaxy density is due to
a cluster at even higher redshift (z $\sim$ 0.95). Therefore \p0537 might
either be projected onto a galaxy cluster at higher redshift or even be part
of a large-scale structure with an extension towards \p0537. 

\section{Summary}

We have presented a deep I-band image of \p0537 as well as multi-object
spectroscopy in the field of \p0537. Our main results are:

1) We could not detect a host galaxy of \p0537 or any galaxy along the
   line-of-sight to the BL Lac down to a detection limit of \mi = 19.0. 
None of the companion
   galaxies to \p0537 is a lensed background galaxy. Thus, we find neither 
evidence that the properties
   of \p0537 are affected by gravitational lensing nor that a potential group
   of galaxies surrounding \p0537 act as a lensing system itself.\\
2) The nearby environment of \p0537 is very complex. The three major findings
   are: i) We detected extended off-nuclear \o2 emission in \p0537. This is
   most likely due to photoionization by the active nucleus, although an
   attractive alternative - jet-cloud interaction by the 
counterjet in \p0537 - can not be
   ruled out. ii) One companion to \p0537 has a very broad, asymmetric
   absorbed  \mg2 line at z = 0.885. This source is most likely a QSO,
   possibly a mini low-ionization BAL QSO. If this can be confirmed, \p0537
   would be the first BL Lac being a member of a binary QSO. 
iii) We further found two companion
   galaxies within a few arcsec at similar redshifts as the BL Lac
(z = 0.892 and 0.895   vs. z = 0.893), which may trigger
   the strong activity in this source. One of the companion galaxies is at the
   location, where the radio jet of \p0537 turns over.\\
3) \p0537 seems to be embedded in a galaxy cluster as rich as Abell type
   0-1. 
This is supported by the detection of four galaxies at similar redshifts 
as the BL Lac ($\Delta$z $<$ 0.002). Since we found serendipitously 
an even higher number of galaxies between z = 0.9 - 1, 
the BL Lac may also be projected onto a galaxy cluster at somewhat higher
redshift (z $\sim$ 0.95) or be part of a large-scale structure with an
extension towards \p0537.

Clearly, the interpretation of the most interesting findings is 
difficult and somewhat speculative. 
High-resolution spectroscopy of the extended \o2 emission line region 
(possibly also in \O3) in \p0537 with the VLT are
required to unambiguously discriminate between the photoionization and
jet-cloud interaction model (e.g. by looking for kinematic signatures 
of a counterjet). A deep high-resolution VLA image of \p0537 would
be useful to search for a counterjet directly. High-resolution spectroscopy
in the blue and moderate-resolution spectroscopy in the 
far-red part of the optical spectrum or even in the NIR is necessary 
to confirm the presence of a (mini low-ionization BAL) QSO and to establish 
the presence of a binary QSO incorporating the BL Lac.

\acknowledgements{We thank the referee, Dr. Elena Pian, for a careful reading
  of the manuscript. It is a pleasure to thank I. Appenzeller for allocating us
GTO-time for the spectroscopic observations at the VLT1. 
We thank also the ESO staff at the VLT and NTT for their
excellent support during the observations, B. Ziegler for taking a short 
acquisition image of \p0537 at the VLT1 for the preparation of the MOS-setups
and P. Cassaro for his VLA 1360 MHz image of \p0537. 
We appreciate suggestions and discussions with M. Dietrich, S. Noll, S. Seitz, 
C. Tadhunter and C. Tapken during the various stages of this project.
This work was supported by the 
Deutsche Forschungsgemeinschaft (SFB 375, SFB 439),
the VW foundation, the German Federal Ministry of Science and Technology
with ID-Nos. 05 2HD50A, 05 2GO20A and 05 2MU104
and the Academy of Finland (project 42697).}

\end{document}